\documentclass[
	twocolumn,
	amsmath,
	amssymb,
	prx,
	aps,
	nofootinbib,showpacs,superscriptaddress,
	]{revtex4-2} 
\usepackage{graphicx}
\usepackage{xcolor}
\usepackage{amsmath}
\usepackage{amssymb}
\usepackage{mathtools}
\usepackage{booktabs}
\usepackage{titlesec}
\usepackage{float}

\begin{document}

\title{Vectorial driving of multistable materials: singularities, pt-graphs, and non-generic paths}

\author{C.M.~Meulblok}
\affiliation{Huygens-Kamerlingh Onnes Lab, Universiteit Leiden, P.O.~Box~9504, NL-2300 RA Leiden, Netherlands}
\affiliation{AMOLF, Science Park 104, 1098 XG Amsterdam, Netherlands}

\author{M.~van Hecke}
\affiliation{Huygens-Kamerlingh Onnes Lab, Universiteit Leiden, P.O.~Box~9504, NL-2300 RA Leiden, Netherlands}
\affiliation{AMOLF, Science Park 104, 1098 XG Amsterdam, Netherlands}

\date{\today}

\begin{abstract}
Describing and predicting the response of multistable materials to external driving is central to memory formation, programmable metamaterials, soft robotics, and in-materia computing. While scalar driving is captured by transition graphs (t-graphs), vectorial driving produces path-dependent responses that require the recently introduced path-transition graphs (pt-graphs). In both cases, transitions are governed by singularities in the energy landscape: for scalar driving these correspond to saddle-node bifurcations, but for vectorial driving, higher-order singularities become important.
Combining experiments on chain-like metamaterials with a minimal spring model, we investigate how higher-order singularities shape pt-graphs and the resulting path-dependent responses. We moreover discuss the role of non-generic driving paths through higher-order singularities, where the response is governed by spontaneous symmetry breaking. 
Finally, we demonstrate how t-graphs emerge as the one-dimensional limit of pt-graphs, unifying scalar and vectorial driving within a common graph-based framework. These results establish a singularity-based approach to path-dependent responses and provide a foundation for designing multistable materials with programmable sequential functionality for smart sensing, soft robotics, and in-materia computation.
\end{abstract}

\maketitle

\section{Introduction}
Multistable materials exhibit intermittent and history-dependent responses, where under sequential driving the system transitions between metastable states \cite{KeimRMP2019,PaulsenAR2025}.
For scalar driving, i.e. using a single control parameter, transition graphs (t-graphs)—which encode stable states as nodes and transitions as directed edges—have emerged as the central framework for understanding multistable materials, revealing and classifying a wealth of memory effects in granular packings, crumpled sheets, disordered magnets, and other frustrated systems\cite{MunganAHP2019,KeimPRR2020,TerziPRE2020,KeimPRR2020,RegevPRE2021,KeimSciAdv2021,LindemanSciAdv2021,LindemanSciAdv2025,BensePNAS2021,ShohatPNAS2022,JulesPRR2022, PaulsenPRSA2019,KeimRMP2019,PaulsenAR2025}.

Many natural and engineered systems, however, are driven by multiple control parameters. Such {\em vectorial driving} gives rise to fundamentally new behavior, in which the trajectory through driving space determines the outcome~\cite{FlorijnPRL2014,SiroteNatComm2024,GuoNat2023,MeulblokArXiv2025}. These encompass non-Abelian behavior, mixed-mode switching, chiral loop transients, emergent computation, and other phenomena with no analogue under scalar driving~\cite{GuoNat2023,SiroteNatComm2024,MeulblokArXiv2025}. To describe these responses, we recently introduced a framework based on path-transition graphs (pt-graphs) for multistable materials under vectorial driving~\cite{MeulblokArXiv2025}.

Here we investigate how singularities shape path-dependent responses~\cite{MuhaxheriNJP2025,YangPNAS2023}. The simplest singularity, the codimension-one fold bifurcation - generate a characteristic pattern consisting of three states—ancestor, descendant, and sibling (ADS)—linked by reversible and irreversible transitions. These ADS triplets form the fundamental building blocks of pt-graphs \cite{MeulblokNonGeneric}.
We probe the role of more complex singularities of higher codimension, using a combination of experiments on chain-like mechanical metamaterials with simulations of an accompanying minimal spring model \cite{MeulblokNonGeneric}.
In particular we show how cusp singularities, pitchfork bifurcations, and a codimension-four butterfly singularity organize the pathways. In particular,
we construct strain maps and pt-graphs that capture the response of multistable systems under generic driving paths and demonstrate how non-generic paths through higher-order singularities can also be understood within the same framework.
Finally, we show how t-graphs emerge as a special one-dimensional limit of pt-graphs, thereby unifying descriptions of scalar and vectorial driving within a common graph-theoretic framework.
Together, our results establish a singularity-based language for path-dependent responses in multistable systems and provide a foundation for designing multistable materials with rich programmable sequential behavior.

\section{System and phenomenology}
We consider chain-like multistable mechanical metamaterials that permit spatially localized compression at multiple sites \cite{MeulblokArXiv2025}. 
Here we briefly recall the system and key observations.

\begin{figure}[t]
	\centering
	\includegraphics[]{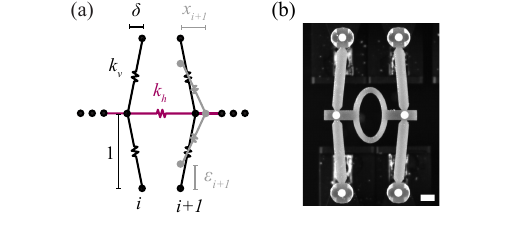}
	\caption{Geometry. 
    (a) Staggered chain of two left-right symmetry broken units (here $i$ and $i+1$) characterized 
    by the dimensionless symmetry breaking $\delta=x_i(\varepsilon_i=0)$ and spring ratio $k:=k_h/k_v$. 
    (b) Top view of the experimental sample A in its rest configuration, where
    the central ellipse materializes a soft horizontal spring. The ends of the vertical beams connect to the strain apparatus   
    (scale bar: 5~mm). 
	}\label{fig:geometry}
\end{figure}

The idealized metamaterial geometry consists of $n$ coupled units, with each unit composed of two identical vertical springs joined at a freely hinging central node and neighboring units linked by horizontal springs.
The 
stiffnesses $k_v$ and $k_h$ of the vertical and horizontal beams define the stiffness ratio $k := k_h/k_v$, and
the 
endpoint separation \(2\ell_0\) sets the length scale 
(Fig.~\ref{fig:geometry}a).
The rest positions of the internal nodes are $x_i = (-1)^{i+1}\delta$, so successive units alternate in pre-curvature direction yielding a nonlinear and deterministic response under compression.
Spatially localized compressive strains $\varepsilon_i$ are imposed to drive this system, and we employ  
driving protocols that are paths $P$ in strain space $\vec{\varepsilon}=
(\varepsilon_1,\varepsilon_2,\dots)$.

Numerically, we 
study a minimal spring model governed by an elastic energy 
\begin{equation}
E = \sum_{i=1}^{n} \left( \frac{x_i^2}{2} - \gamma_i \right)^2 
+ \sum_{i=1}^{n-1} \frac{k}{2} \left( x_i - x_{i+1} - (-1)^i2\delta \right)^2 ,
\label{eq:energy}
\end{equation}
where $\gamma_i := \varepsilon_i + \delta^2/2 - \varepsilon_i^2/2 \approx \varepsilon_i + \delta^2/2$.
For limit small $\delta$, a rescaling of variables eliminates $\delta$, leaving the effective interaction strength $k/\delta^2$ as the key system-dependent parameter \cite{MeulblokArXiv2025} (see appendix~B).

We materialize such metamaterials experimentally, in 
samples fabricated from flexible silicone rubber (see appendix~A). 
Spatially localized compressive strains $\varepsilon_i$ are imposed using a custom-built device that symmetrically shortens the vertical distance between the endpoints of each unit, $2\ell = 2\ell_0(1-\varepsilon_i)$, while maintaining horizontal alignment of all nodes. 
We focus on two samples with distinct path dependencies and
structure of their singularities:
sample A with $(\delta, k)\approx(0.08, 0.09)$ and sample B with $(\delta, k)\approx(0.08, 0.04)$
(Fig.~\ref{fig:geometry}b; see appendix~A and \cite{MeulblokArXiv2025}).

\subsection{ADS-triplets} \label{sec:PDs}

\begin{figure}[h!]
    \centering
    \includegraphics[]{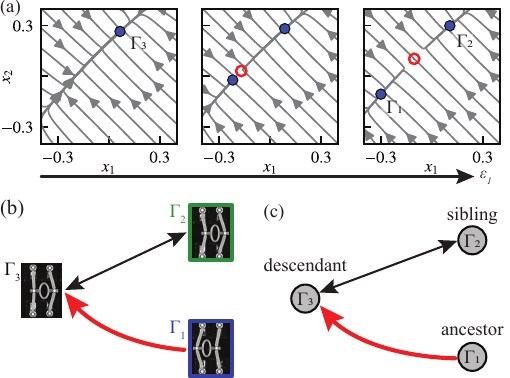}
    \caption{Fold bifurcation and ADS-triplet.
    (a) Theoretical evolution of the gradient for $(\delta,k)=(0.1, 0.09)$ (sample A), with $\varepsilon_2=0.03$ and $\varepsilon_1=0,~0.016,~0.03$ (left to right). Markers denote stable (blue) and unstable (red) fixed points. 
    (b) Real-space evolution of sample A under the same driving protocol, showing configurations at $\vec{\varepsilon}=(0,0.03)$ (left) and $(0.03,0.03)$ (right).
    (c) Graph representation of the general structure underlying path dependency: three configurations--ancestor ($\Gamma_1$), descendant ($\Gamma_3$), and sibling ($\Gamma_2$)--connected by reversible (black) and irreversible (red) transitions.
    }\label{fig:fold}
\end{figure}

Under vectorial driving, two-unit systems can exhibit a non-Abelian response—where configurations depend on the order of straining beams 1 and 2—and a mixed-mode response, where smoothness of the response depends on the driving path but not on the initial or final strains \cite{MeulblokArXiv2025}. 
Discontinuities of the response occur at fold (saddle–node) bifurcations (Fig.~\ref{fig:fold}a) \cite{MeulblokArXiv2025,MuhaxheriNJP2025}.
Each fold bifurcation naturally organizes three configurations—ancestor, descendant, and sibling—linked by an irreversible transition from ancestor to descendant and a reversible continuation between descendant and sibling (Fig.~\ref{fig:fold}b-c).

\section{General framework: pt-graphs}

In this section, we develop a general framework to describe the path-dependent response of multistable systems under vectorial driving.
First, we introduce path-transition graphs (pt-graphs) that are able to capture the response of multistable systems for generic strain paths. 
We illustrate this approach for two-beam examples in two parameter regimes (Sec.~\ref{sec:ptgraphs}). 
We then discuss the behavior along non-generic strain paths, such as paths through higher-order singularities, and argue that nearby paths are captured by pt-graphs as well as specific non-generic paths (Sec.~\ref{sec:nong}).
We close this section by
discussing the connections between
t-graphs and pt-graphs (Sec.~\ref{sec:1dvs2d}).

\subsection{Strain Maps and PT-Graphs} \label{sec:ptgraphs}

We now introduce pt-graphs to characterize the sequential response of multistable systems to any (generic) strain path. In these graphs, the nodes represent states, and their connections are a mix of undirected and directed edges, corresponding to smooth and irreversible transitions respectively \cite{MeulblokArXiv2025}.

We start by identifying an initial configuration - customarily, one that is stable at zero strain - and then use a recursive algorithm to determine the states, strain-map and pt-graph.
Starting from the initial configuration, we follow its evolution along paths throughout the strain domain
 and keep track of the irreversible transitions and corresponding new states that emerge through ancestor-descendant bifurcations -
thereby where these irreversible transitions occur.
Note that in $d$ dimensional strain-space, the fold locations lie on $d-1$ dimensional hypersurfaces.
Then, we repeat this procedure for all new configurations. This yields the strain map, which identifies and labels the locus of irreversible transitions. The strain map  
partitions the strain-space into domains. States 
are defined as equivalence classes of configurations that are smoothly connected within each domain. 

After the strain map and states are identified, we collect this information in the pt-graph. The states form the nodes of the pt-graph, while 
the reversible and irreversible transitions which occur when the driving crosses a boundary in the strain map form the undirected and directed edges of the pt-graph. The result is a mixed graph, where all edges are labeled by the corresponding boundary of the strain map. The combination of the strain map and the pt-graph then specifies the evolution.

\subsubsection{Examples of pt-graphs}
To illustrate the procedure of obtaining the strain map and pt-graph, we experimentally determine these for
samples A and B, which differ in their stiffness ratio $k$ (see appendix~A).

\begin{figure}[tb]
\centering
\includegraphics[]{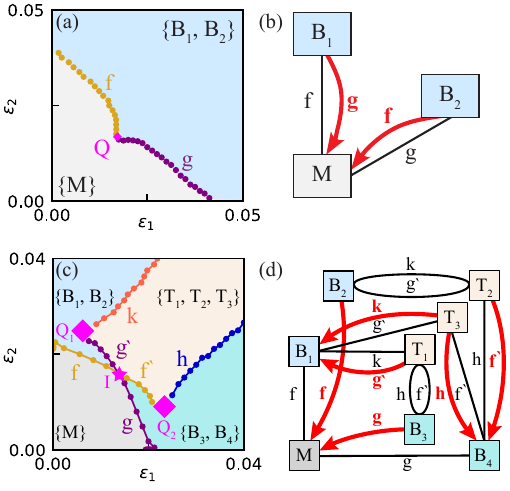}
    \caption{Strain maps and path-transition (pt) graphs for two-unit systems.
    (a-b) Sample A ($k\approx 0.09$).
    (a) Strain map showing the organization of fold singularities. The codimension-one fold curves $f$ and $g$ partition strain space into a monostable region with state $M$ (gray) and a bistable region with states $B_1$ and $B_2$ (blue). The curves meet at point $Q$, corresponding to a codimension-two cusp singularity.
    (b) Corresponding pt-graph encoding the response: nodes represent states, with reversible connections shown as undirected black edges and irreversible transitions across fold curves as directed red edges.
    (c,d) Sample B ($k\approx 0.04$).
    (c) Strain map with a more complex arrangement of fold curves, including cusp singularities $Q_1$ and $Q_2$ (pink squares). The intersection point $I$ (pink star) marks where distinct fold branches cross, requiring a distinction between transition pathways ($f$ vs.\ $f'f$, and $g$ vs.\ $g'$).
    (d) Corresponding pt-graph, showing the enriched path-dependent structure arising from the additional singularities.
	}\label{fig:ptgraph}
\end{figure}

{\em Sample A---}These is only one stable configuration at zero strain. By following its evolution under clockwise and counter clockwise strain loops (see appendix~A), we find two distinct transition curves labeled $f$ and $g$ that meet in a point labeled $Q$; these curves  specify the location of fold singularities in strain space \cite{MeulblokArXiv2025}
(Fig.~\ref{fig:ptgraph}).

The fold curves $f$ and $g$
partition the strain space into a monostable and a bistable domain
(Fig.~\ref{fig:ptgraph}a).
In the monostable domain, the evolution is smooth and reversible, and all configurations are thus equivalent and are represented by 
a single state that we label \(M\)
(Fig.~\ref{fig:ptgraph}a).
At each point in the bistable domain, there are two 
stable configurations which smoothly deform along strain paths within the bistable domain, thus defining two states $B_1$ and $B_2$. 
Note that
configurations associated with states $B_1$ and $B_2$ cannot smoothly deform into each other for strain paths {\em within} the bistable domain. 
Hence, determining the locus of the fold bifurcations identifies the states and strain map (Fig.~\ref{fig:ptgraph}a).

To create the corresponding pt-graph, we 
determine how these states are connected when the strain crosses the fold curves (for now we ignore singular paths through $Q$). State $M$ smoothly transforms into state $B_1$ when we cross from the monostable to the bistable domain across $f$, and smoothly transforms into state $B_2$ across $g$; these smooth paths are reversible, so that $B_1$ and $B_2$ smoothly connect to $M$ across $f$ and $g$, respectively. The irreversible transitions occur when, starting from $B_1$, we cross boundary $g$ from the bistable to the monostable regime --- across this boundary, we observe the irreversible transition $B_1\!\rightarrow\! M$; similarly, across
$f$  we observe the irreversible transition $B_2\!\rightarrow\! M$. We summarize these transitions in a mixed graph with three
nodes for the states $M$, $B_1$ and $B_2$, which are connected by directed and undirected edges corresponding to irreversible and reversible transitions across $f$ and $g$  (Fig.~\ref{fig:ptgraph}). 

This pt-graph illustrates two general properties. First, each fold curve is associated with an ADS-triplet.
Here, the ancestor, descendant and sibling associated with curve $f$
are
$(B_2,M,B_1)$,
which we denote as $ADS_f=(B_2,M,B_1)$; similarly,
$ADS_g=(B_1,M,B_2)$.
Second, the topology of the strain map constrains the number of edges in the pt-graph, as each state and each adjacent fold-boundary correspond to an outgoing edge in the pt-graph. The number of outgoing edges is $\Sigma_{domain} n_s n_f $, where $n_s$  and $n_f$ are the number of states and adjacent fold curves per domain; here $\Sigma_{domain} n_s n_f =6$. As reversible edges
in the pt-graph correspond to a pair of incoming and outgoing edges, this implies that
$\Sigma_{domain} n_s n_f =  2n_r + n_i$, where
$n_r$ and $n_i$ are the numbers of reversible and irreversible edges of the pt-graph.

{\em Sample B---}To demonstrate the generality of this approach,
we apply our framework to characterize sample B which has weaker
coupling ($k\approx 0.04$) and a more complex pt-graph
(appendix~A).
Starting from the neutral configuration at zero strain, we perform
the recursive algorithm for strains $0\le \varepsilon_i \le 0.04$ and find that we require four families of strain loops to determine  the strain map (appendix~A).
The strain-space is partitioned by the fold curves
($f,~f',~g,~g',~h,~k$) into four domains: a monostable domain containing state $M$, two bistable domains with states $B_1$ - $B_4$, and one tristable domain containing states $T_1$, $T_2$ and $T_3$
--- hence
$\Sigma_{domain} n_s n_f =26$
(Fig.~\ref{fig:ptgraph}c).

The corresponding pt-graph features eight states and
has 16 edges ($n_i=6$, $n_r=10$), consistent with the relation
$\Sigma_{domain} n_s n_f =  2n_r + n_i$ (Fig.~\ref{fig:ptgraph}d).
Each of the six boundaries corresponds to
an ADS triplet:  
$ADS_f=(B_2,M,B_1)$,
$ADS_g=(B_3,M,B_4)$,
$ADS_{f'}=(T_2,B_4,T_3)$,
$ADS_{g'}=(T_1,B_1,T_3)$,
$ADS_k=(T_3,B_1,T_1)$, and
$ADS_h=(T_3,B_4,T_2)$.
In addition, there are
four reversible pairs across the boundaries of the tristable domain:
$RP_{f'}=(T_1,B_3)$,
$RP_{g'}=(T_2,B_2)$,
$RP_k=(T_2,B_2)$,
$RP_h=(T_1,B_3)$.
These correspond to configurations that are smoothly connected. However, as discussed above, we associate these with distinct states, as their respective transitions across other boundaries produce different states: e.g., 
even though $T_1$ and $B_3$ are reversibly connected across $f'$, a path from $B_3$ over $g$ to $M$ produces an irreversible transition, while a path from $T_1$ over $k$ and $f$ smoothly transitions to $M$.  
This more involved example illustrates that each pt-graph is constructed from a combination of ADS triplets, one for each section of a fold curve, and reversible pairs, one for each state not involved in the ADS triplet across each section of a fold curve. 

\subsubsection{Subtleties}
The combination of strain map and pt-graph gives a powerful tool to
characterize and distinguish path dependencies in multistable systems. However, there are two subtleties. 

First, as path dependencies lead to non-local properties, 
extending the strain range of the strain map 
may lead to new states and transitions even in the regime that was already mapped out previously. To illustrate this, we consider 
what happens with the strain maps of sample A in two strain ranges.
In the large strain range  ($\varepsilon_1 \in [0,0.05], \varepsilon_2 \in [0, 0.05]$) we find 
$Q$, $f$ and $g$.
Restricting these transitions to a 
small
strain range ($\varepsilon_1 \in [0,0.015], \varepsilon_2 \in [0.02, 0.04]$) would appear to only select transition $f$. 
However, to detect $f$, we need the consider ancestor state $B_2$. 
But if we start from state $M$ and only consider paths in the small strain range, $B_2$ is never reached, as this requires paths crossing $g$. Hence, for paths in the small strain range, there is only one state, $M$, and there are no transitions.
We note that the strain map and pt-graph for small strain range direct follow from those of the large range: if we remove all transitions along $g$ in the pt-graph, $B_2$ is no longer accessible from $M$, and we are left with reversibly linked pair of states $B_1$ and $M$; we then realize that $f$ can only be detected from $B_2$, and thus in the restricted range, configurations originally associated with $M$ and $B_1$ now are associated with a single state.

The same issue also arises for sample $B$, if we, e.g., consider a restricted strain range that contains $f$, $'f'$, $g$, $g'$ and $I$, but none of the cusp singularities; again, the ancestors of these transitions are not reachable, so that within the restricted strain ranges the fold curves are not detected, and the restricted strain map and pt-graph are trivial (one state, no transitions). 
We stress again that this subtlety arises from the complex non-local definition of a state, and are an unavoidable feature of pathway dependent responses.

A second subtlety may arise when we consider transitions between domains that both contain multiple state. If crossing this border, one configuration undergoes a discontinuous transition, it is not a priory clear in what stable state it will end up, so it is not clear which descendant belongs to this ancestors. In physical systems such ambiguities are resolved by a dynamical rule that stipulates the dynamics of an unstable state --- knowledge of the static energy landscape and its minima is in general not sufficient to predict where an unstable configuration evolves to \cite{jinPRL2025}. We note that this scenario does not arise in the data that we present here.


\subsection{Non-generic paths and higher order singularities}\label{sec:nong}

Generic paths only cross fold singularities, but 
in higher dimensional strain spaces, one generically expect
more complex singularities. For example, for sample A, $f$ and $g$ meet in a singular point $Q$, which we show below to be a co-dimension two cusp singularity. While fold bifurcations govern the ADS patterns, their global organization is dictated by higher-codimension singularities which in turn organize the path-dependent response.

First, intersections of fold lines gives rise to non-generic paths. We demonstrate that such paths are well described by fold singularities and ADS-triplets, and are fully captured in the pt-graph framework (section.~\ref{ss:cross}).
Second, fold lines can meet in co-dimension two cusp singularity.
 We show that paths passing through a cusp are associated with a pitchfork bifurcation. Here the response is 
non-deterministic, but as any regular path nearby is still described by fold bifurcations, the pt-graph description remains valid (section.~\ref{ss:cusp}). 
Third, even higher order singularities arise when we consider, e.g.,
the response of two-beam systems as function of the coupling strength $k$. In particular, the distinct strain maps of samples A and B, which reflect differences in the organization of the fold singularities in the strain space, stem from a co-dimension four butterfly singularity in the $(\vec{\varepsilon},k)$-space
(section.~\ref{ss:hos}).

\subsubsection{Intersections of fold lines}\label{ss:cross}
\begin{figure}[tb]
    \centering
    \includegraphics[]{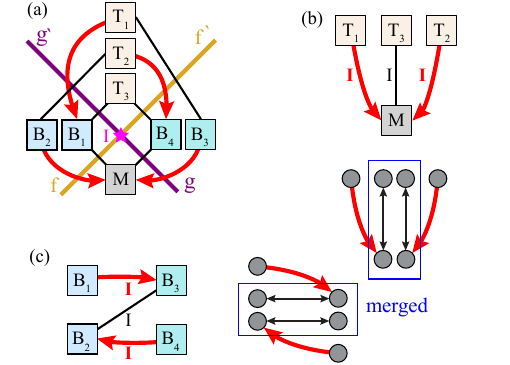}
    \caption{Non-generic paths through intersection point $I$ of sample B.
    (a) Part of the pt-graph showing the transitions near the intersection point $I$. (b) The transitions from the tristable domain to the monstable domain through $I$ are unambiguously determined by the pt-graph (a). Note that the transitions can be described by two merged ADS patterns.
    (c) The transitions between the two bistable domains through $I$ are unambiguously determined by the pt-graph (a). Note that the transitions can be described by two merged ADS patterns.
    }\label{fig:NonGenericCrossing}
\end{figure}

To illustrate that the pt-graph fully describes strain path that cross through intersections of fold bifurcations, we
consider sample B, subject to strain paths through intersection point $I$ (Fig.~\ref{fig:ptgraph}c).
For each path linking the monostable state $M$ to one of the three tristable states $T_i$, the 'aggregate' transition can be unambiguously classified as reversible or irreversible (Fig.~\ref{fig:NonGenericCrossing}a). In particular, the corresponding ADS patterns can be consistently merged to determine the net response for paths traversing $I$ (Fig.~\ref{fig:NonGenericCrossing}b). 
A similar construction applies for paths connecting the two bistable domains: their aggregate transitions are again uniquely classified as reversible or irreversible, and the associated ADS triplets combine to yield the response across $I$ (Fig.~\ref{fig:NonGenericCrossing}c). Hence, despite their non-generic nature, paths passing through such intersection points remain unambiguously described within the pt-graph framework.

\subsubsection{Cusp singularities} \label{ss:cusp}
\begin{figure}[tb]
\centering
\includegraphics[width=\linewidth]{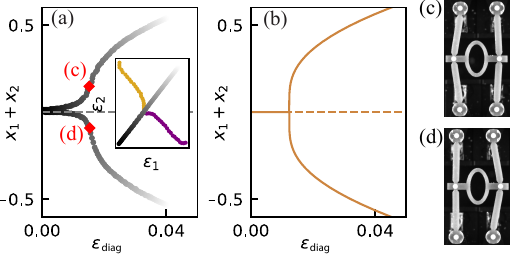}
\caption{
Pitchfork bifurcation observed along a non-generic strain path through a cusp singularity. (a) The collective coordinate $x_1 + x_2$  along a diagonal strain path  ($\varepsilon_1=\varepsilon_2:=\varepsilon_{\text{diag}}$, inset)
starting from $\varepsilon_{\text{diag}}=0.04$ and ending at $\varepsilon_{\text{diag}}=0$ (sample A).
(b) Results from spring model ($k=0.09$ and $\delta=0.08$) along the same path. (c-d) Snapshots of the system corresponding to the red dots indicated in panel (a); note that in both cases, the left beam is left leaning and the right beam is right leaning. 
	}\label{fig:NonGenericType2}
\end{figure}

Next, we consider strain paths through the cusp singularities in the strain maps which correspond to pitchfork bifurcations (Fig.~\ref{fig:ptgraph}).
To do so, we revisit sample A and consider a diagonal strain path through the cusp singularity at point $Q$, defined by $\varepsilon_{\mathrm{diag}}\equiv \varepsilon_1=\varepsilon_2$. We track the evolution of the bistable states as 
$\varepsilon_{\mathrm{diag}}$ is decreased from $0.04$ to $0$, complemented by numerical analysis (Fig.~\ref{fig:NonGenericType2}a–b).
In the experimental system, we observe a near-perfect pitchfork bifurcation, with two symmetry-related branches (Fig.~\ref{fig:NonGenericType2}c-d); 
a similar pitchfork bifurcation is observed for the cusp singularities of sample B.
In the numerical model, paths through the cusps correspond to a perfect pitchfork bifurcation (Fig.~\ref{fig:NonGenericType2}b).

Accessing such singular points requires tuning two independent parameters, and cusps therefore occur generically only in higher-dimensional strain spaces.
Along the diagonal path from $\varepsilon_{\text{diag}}=0$ to $0.04$, the pitchfork bifurcation induces a non-deterministic evolution. Therefore, the evolution along such a path is not strictly captured by the pt-graph framework; however, the pt-graph perfectly captures the potential transitions that occur of generic pathways close to this singular path.
This is not dissimilar from specific one-dimensional strain paths that just reach, but do not cross, a switching threshold for scalar driving described by a t-graph.

\subsubsection{Butterfly singularity: the transition between strong and weak coupling}
\label{ss:hos}

Higher order singularities may also arise when we consider the path-dependent response of a family of systems; for example, 
the strain maps of samples A and B reflect different organizations of fold singularities in strain space and originate from a co-dimension four butterfly singularity in $(\vec{\varepsilon},k)$-space.

\begin{figure}[tb]
    \centering
    \includegraphics[]{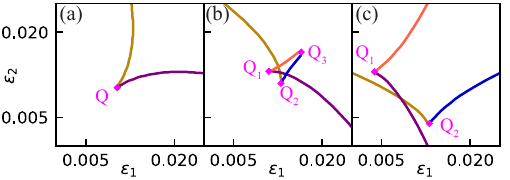}
    \caption{Evolution of strain map as a function of coupling $k$.
    (a-c) Numerically obtained strain map for two-unit systems with $\delta=0.09$, and $k=0.2$, $0.06$, and $0.02$ respectively.
    }\label{fig:CouplingRegimes}
\end{figure}

We numerically track the strain map as a function of the coupling coefficient $k$ for fixed $\delta$ - recall that the effective interaction scales as $k/\delta^2$. We
identify a critical coupling $k^*$ where the system exhibits a butterfly singularity, which separates a `strong' and `weak' coupling regime 
(Fig.~\ref{fig:CouplingRegimes}).
For strong coupling ($k>k^*$), the strain map exhibits a single cusp singularity and a supercritical pitchfork bifurcation along the diagonal strain path $\varepsilon_{\mathrm{diag}}$, as in sample A (Figs.~\ref{fig:NonGenericType2} and \ref{fig:CouplingRegimes}a). 
Lowering the coupling, we reach a butterfly singularity at $k=k^*$. We note that we can reach this codimension-four singularity by tuning only three parameters due to the left-right symmetry of the two-unit system. 
Just below $k^*$, the butterfly unfolds into three cusps ($Q_1\!-\!Q_3$), constituting the canonical unfolding of a codimension-four butterfly singularity \cite{Arnold2003} (Fig.~\ref{fig:CouplingRegimes}b).
Concomitantly, the pitchfork along $\varepsilon_{\mathrm{diag}}$ changes from supercritical ($k>k^*$) to subcritical ($k<k^*$).
As $k$ decreases further into the weak coupling regime ($k<k^*$), the cusps $Q_1$ and $Q_2$ approach $\varepsilon_1=0$ and $\varepsilon_2=0$ along the lines $\varepsilon_2\approx3\delta^2/2$ and $\varepsilon_1\approx3\delta^2/2$ respectively (see appendix~B).
In addition, $Q_3$ moves to infinity, leaving two cusps as observed in  sample B (Fig.~\ref{fig:CouplingRegimes}c).
We numerically determined the critical coupling $k^*/\delta^2 \approx 10$ (see appendix~B). 

The emergence of the butterfly singularity accounts for the increased complexity of responses in the weak coupling regime, directly linking the singularity structure to the path dependent responses of samples A and B.

\subsection{PT-Graphs vs T-Graphs} \label{sec:1dvs2d}

Scalar driving is a special case of vectorial driving. Here, we discuss  the relation between our pt-graph framework and the t-graphs developed to describe the sequential response under scalar driving
\cite{RegevPRE2021,LindemanSciAdv2021,KeimSciAdv2021,LiuPNAS2024,MunganPRL2019,PaulsenPRSA2019,MelanconAdvFuncMat2022,DingJCP2022,TeunisseArXiv2024}.

The nodes in a t-graph 
represent states which are equivalence classes of configurations: two configurations correspond to the same state if they smoothly deform into each other under slow increase or decrease of the driving strain $U$. 
The nodes are connected by directed edges that represent irreversible transitions. Each transition is characterized by, first, whether it is an up or down transition, i.e. triggered by increasing or decreasing $U$, and second, the 
switching thresholds $U^\pm(S)$  \cite{BensePNAS2021,KeimSciAdv2021,LiuPNAS2024,PaulsenPRSA2019,HeckePRE2021,TeunisseArXiv2024}. The t-graph with labeled edges
specifies the response under any (generic) quasistatic driving protocol.

There are striking similarities between pt-graphs and t-graphs.
Both combine a graph and a strain map - for t-graphs the strain map
is simply given by the collection of switching thresholds \cite{LiuPNAS2024,TeunisseArXiv2024}. 
Their nodes represent states, and their edges correspond to transitions that occur at a codimension-one subspace of the driving-space: points given by switching thresholds in the case of scalar driving, and fold manifolds for vectorial driving. 

However, there are also striking differences.
Most importantly is the subtle difference in the definition of what constitutes a state. For vectorial driving, two states can be connected by both a reversible and irreversible transition, associated with two different fold singularities; for scalar driving, this never occurs. Hence, while for scalar driving the group of configurations connected by a smooth deformation form one state, for vectorial driving we define states as groups of configurations that smoothly deform into each other for 
deformation paths that do not cross any fold transition
(see Sec.~\ref{sec:ptgraphs}). 
As a consequence, t-graphs only contain irreversible transitions,
whereas pt-graphs contain both reversible and irreversible transitions (Fig.~\ref{fig:ptgraph}). 

\begin{figure}[t]
	\centering
	\includegraphics[]{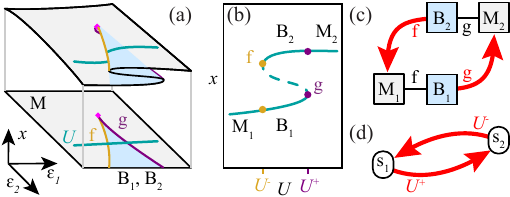}
	\caption{The relation between pt-graphs and t-graphs. 
    (a) Schematic representation of the configuration space near the cusp singularity of sample A, with a one-dimensional path parametrized by $U$  (green). 
    (b) The collective coordinate $x$ as a function of the scalar driving $U$.
    (c) Along the one-dimensional path, we can construct a pt-graph.
    (d) By identifying uniquely, smoothly connected states, we obtain a standard t-graph, with two states and edges labelled by the up- and down-switching thresholds $U^\pm$.
    }\label{fig:1dvs2d}
\end{figure}

We now show that t-graphs are a special case of pt-graphs. 
We illustrate the mapping from {a} pt-graph with one-dimensional driving to {a} t-graph
by using the example of a cusp singularity as displayed by sample A. Consider the pt-graph along a one-dimensional curve, so that $\vec{\varepsilon}$ is a function of the scalar driving $U$ (Fig.~\ref{fig:1dvs2d}a). As {a} function of $U$, the collective coordinate $x:=x_1+x_2$  forms an elementary hysteresis loop, {constructed} by appropriately connecting two fold bifurcations (Fig.~\ref{fig:1dvs2d}b).
The fold bifurcations $f$ and $g$ occur at switching thresholds $U^-$ and $U^+$, and 
are associated with ADS-triplets: $ADS_f=(B_2,M_{{1}},B_1)$ and $ADF_g=(B_1,M_{{2}},B_2)$. Note that while $M_1$ and $M_2$ are the same state in the pt-graph for two dimensional driving (Fig.~\ref{fig:ptgraph}), as there is a smooth path that connects them that does not cross $f$ or $g$, in the pt-graph for one dimensional driving, $M_1$ and $M_2$ are not smoothly connected along $\vec{\varepsilon}(U)$ and form  distinct states.
Hence, restricted to the one-dimensional path, we obtain a pt-graph with four states, two irreversible edges and two reversible edges (Fig.~\ref{fig:1dvs2d}c). 

The pt-graph for one dimensional driving can be simplified further by clustering
smoothly connected pairs of states in one-dimensional pt-graphs
-- here $(M_1,B_1)$ and
$(M_2,B_2)$ ---  as states $S_1$ and $S_2$ {(Fig.~\ref{fig:1dvs2d}c-d)}.
{This procedure is well-defined since}
the location of the fold singularities 
form points on a line and are well-ordered; hence states that are
connected smoothly cannot be connected irreversibly through a different path.
In our example, this clustering results in a t-graph with two states and only irreversible transitions (Fig.~\ref{fig:1dvs2d}d). For this example, the t-graph represents a single elementary hysteresis loop. More generally, 
for any pt-graph that is restricted to a one-dimensional driving subspace, a similar mapping to t-graphs can be constructed: t-graphs are thus special cases of pt-graphs -  and  pt-graphs
generally encode a large number of t-graphs that can be found by restricting the strain space to one-dimensional manifolds.

\section{Conclusion}
We have developed a general framework for describing path-dependent responses in multistable systems under arbitrary vectorial driving. By combining experiments and numerical simulations, we demonstrated that the complex behavior induced by multidimensional control parameters can be systematically understood in terms of singularities of the energy landscape and their graph-based representation through path-transition graphs. ADS triplets are the elementary building blocks of path-dependent responses. Fold bifurcations generate these triplets, while higher-order singularities organize them into distinct classes of strain maps and responses, with cusp singularities producing pitchfork bifurcations and a codimension-four butterfly marking the boundaries between response classes.

Our pt-graph framework extends the t-graph description of scalar driving to multidimensional driving spaces, suggesting that ADS triplets play a role analogous to that of hysterons in theories of memory and hysteresis under scalar driving \cite{RegevPRE2021,LindemanSciAdv2021,KeimSciAdv2021,LiuPNAS2024,MunganPRL2019,PaulsenPRSA2019,MelanconAdvFuncMat2022,DingJCP2022,BensePNAS2021,ShohatPNAS2022,MunganAHP2019,TerziPRE2020,HeckePRE2021,JulesPRR2022}. 
Finally, by establishing a direct link between multistability, path dependence, and singularities, our framework transforms path dependence from a challenge into a design resource for smart sensing, soft robotics, adaptive metamaterials, and in-materia computation.
 

\cleardoublepage

\appendix

\section{Experimental details} \label{sec:expdetails}
\subsection{Geometry and Fabrication}
Experimentally, we focus on two distinct samples that differ only in the geometry of the horizontal beams resulting in distinct effective coupling $k$.
The vertical beams in each geometry have thickness 3.5 mm and have a tapered connection to minimize torsional effects (Fig.~\ref{fig:geometry}a).
All our samples have a thickness of $10~\text{mm}$ to prevent out-of-plane buckling.
We embedded an elliptical annulus in the horizontal beam to lower the effective spring constant. The precise geometry of the elliptical annulus differs for each sample (Fig.~\ref{fig:geometry}b, Tab.~\ref{tab:horizontalspring}) (See Sec.~\ref{sec:kstar} for $k$ measurements).
Finally, we design $2.5~\text{mm}$ diameter protrusions on each node to measure the real-space configuration (Sec.~\ref{sec:realspace}).

The physical samples are fabricated by pouring degassed
Mold Star 30 silicone rubber
(Young's modulus of $\approx1$ MPa, Poisson's ratio of $\approx 0.5$)
into open face molds that are 3D printed on UltiMaker S3 printers.
The samples are cured at room temperature. After curing, the samples are removed by breaking the molds carefully.

\begin{figure}[ht]
	\centering
	\includegraphics[]{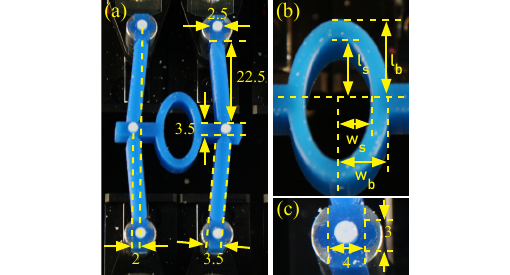}
	\caption{Sample geometry; all dimensions in milimeters, errorbar   $\pm 0.1~\text{mm}$. (a) Sample A with two units,
   (b) Zoom in on elliptical annulus. (c) Zoom in on terminating block.
	}\label{fig:geometry}
\end{figure}

\begin{table}[htb]
	\begin{tabular}{c|c|c|c|c}
			 & $w_b$ (mm) & $l_b$ (mm) & $w_s$ (mm) & $l_b$ (mm) \\ \hline
	Sample A & 12    & 20    & 8     & 16    \\ \hline
	Sample B & 12    & 21    & 9     & 18
	\end{tabular}
	\caption{Specification of the horizontal beam for each sample, each measurement has an error of $\pm 0.1~\text{mm}$.
	}\label{tab:horizontalspring}
\end{table}

\subsection{Measuring Effective Coupling}\label{sec:kstar}
\begin{figure}[hbt]
    \centering
    \includegraphics[]{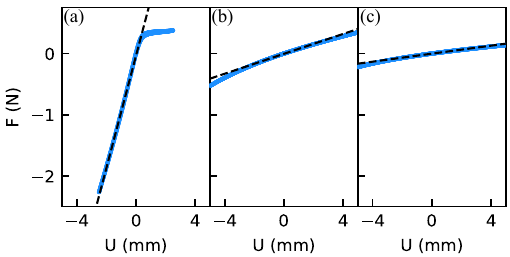}
    \caption{ (a-c) Force-extension curves of the vertical spring, horizontal spring of sample A, and horizontal spring of sample B, respectively. We fit a spring constant in the linear response (black dashed line) and find $k_v\approx9.2\cdot10^2~\text{N/M}$ (a), $k_{h, A}\approx81~\text{N/M}$ (b), and $k_{h, B}\approx33~\text{N/M}$ (c). As a result, the effective coupling of sample A $k_A=0.089 \pm 0.001$ and for sample B $k_B=0.036 \pm 0.001$.
    }\label{fig:kstar}
\end{figure}

We measure the mechanical response of the vertical and both horizontal springs to determine the effective coupling of sample A $k_A=k_{h,A}/k_v$ and sample B $k_B=k_{h,B}/k_v$ (the ratio between the horizontal and vertical spring constants).
We fabricate both springs disconnected from the system using the same 3d-printing and molding techniques, and then probe their mechanical response in an uniaxial testing device (Instron type 5965), which controls the compression $U$ better than $10^{-3}$ mm and allows us to measure the compressive force $F$ with an accuracy $10^{-4}$ N.
Firstly, we present the force $F(U)$ for the vertical spring in Fig.~\ref{fig:kstar}a, and fit the linear response in the domain $U\in[-2~\text{mm},~0.5~\text{mm}]$ and obtain $k_v=0.917\pm0.001 \cdot 10^3$ N/m.
Note that the for large compression ($U>0.7~\text{mm}$), the vertical spring buckles, hence $k_v$ drastically decreases.
However, in our path dependent experiments, we do not reach this buckling limit and therefore observe minimal effects  the path dependent response.
Secondly, we present the force $F(U)$ of the horizontal spring for sample A and B in Fig~\ref{fig:kstar}b and Fig~\ref{fig:kstar}c, respectively.
We fit the linear response in the domain $U\in[-2~\text{mm},~2~\text{mm}]$ and find effective coupling $k_A=0.089\pm0.001$ and $k_B=0.036\pm 0.001$.


\subsection{Driving} \label{sec:realspace}

We use a custom-made actuation device to independently compress each beam. Each beam is connected at both ends to a DC servo motor actuator (Thorlabs Z825B).
To ensure quasi-static driving conditions, the motors operate at a constant velocity of $0.1$ mm/s (corresponding to a strain rate of $5\times10^{-4}\,\mathrm{s}^{-1}$); accelerations are set to $3$ mm/s.
Each motor has a positional accuracy and repeatability of $0.01$ mm, yielding a strain accuracy of approximately $10^{-3}$.
During actuation, we capture snapshots of the configuration using a CCD camera (basler acA2040-90um, $12.9$ pixel/mm). From the snapshots, the node positions are extracted using standard python image analysis packages, achieving resulting an accuracy of approximately $0.005$ in $\{x_i\}$.

\subsubsection{Driving protocols} \label{sec:Paths}
Here, we discuss the families of strain paths used to map the loci of fold curves for sample A and B.
These strain paths consists of a compression step $c_i$, that increases $\varepsilon_i$ from a minimum value $\varepsilon_i^{m}$ to a maximum value $\varepsilon_i^{M}$ while the other strains remain fixed;
and a decompression step $d_i$ reverses this step.
$c_{ij}$ indicates that both beam $i$ and $j$ are compressed simultaneously.
Moreover, in many experiments we take $\varepsilon_i^{m}=0$ and $\varepsilon_i^{M}=\varepsilon^{M}$, so that $\varepsilon^{M}$ sets the overall strain scale. Sequences of operations are denoted using square brackets; for example $[c_1,c_2,d_1,d_2]$ denotes a loop-like compression protocol.

For sample A, we employ two loop-like driving protocols $P_1 : [c_1, c_2, d_1, d_2]$, and $P_2 : [c_2, c_1, d_2, d_1]$, where in $P_1$ ($P_2$) we fix  $\varepsilon_{1}^M=0.05$
($\varepsilon_{2}^M=0.05$) and sweep $\varepsilon_{2}^M$ ($\varepsilon_{1}^M$). Using $P_1$ ($P_2$), we obtain fold curve $f$ ($g$).

For sample B, we apply both rectangular protocols $P_1$ and $P_2$, and triangular protocols $P_3 : [c_{12}, d_1, d_2]$ and $P_4 : [c_{12}, d_2, d_1]$. For $P_1$ and $P_2$, we use the same parameter ranges as in the characterizition of sample A. For $P_3$ and $P_4$, we set $\varepsilon^M_1=\varepsilon^M_2=\varepsilon^M$ and sweep $\varepsilon^M\in[0, 0.04]$.
Using $P_1$ ($P_2$), we obtain fold curves $f$ and $f'$ ($g$ and $g'$). The fold curves $f$ and $f'$ ($g$ and $g'$) correspond to the same irreversible transition, but are partitioned as they connect two distinct boundaries (Fig.~\ref{fig:ptgraph}c).
Using $P_3$ ($P_4$), we obtain fold curve $k$ ($h$).

\section{Numerical Details}
Complementary to experiments, we use a numerical spring model for a systematic in-dept study of path dependency.
We define the elastic energy of our linear spring model $E(x_i)$: In this model we approximate each biased beam with two (nearly) vertical springs with spring constant $k_v$ and rest length $l_0=\sqrt{1+\delta^2}$, where $\delta$ is the bias of the beam.
Thus the energy of each vertical spring is:
\begin{align}
	E_{v_i} = \frac{1}{2}(l-l_0)^2 = \frac{1}{2} \left( \sqrt{ (1-\varepsilon_i)^2 + x_i^2 } - \sqrt{1+\delta^2}\right)^2~.
\end{align}
Here, $x_i$ is the position of the middle of the beam with respect to the clamped-clamped boundaries and $2\varepsilon_i$ is the compressive strain.
The horizontal spring has alternating rest length $1-(-1)^i2\delta$ and spring constant $k_h$, thus the energy of each horizontal spring is:
\begin{align}
	E_{h,i} = \frac{k}{2}(x_i - x_{i+1} - 2\delta)^2 ~.
\end{align}
As each beam in our model system consists of two vertical springs, we obtain a total energy:
\begin{multline}
	E = \sum_{i=1}^{n} \left( \sqrt{ (1-\varepsilon_i)^2 + x_i^2 } - \sqrt{1+\delta^2} \right)^2 + \dots \\
	\sum_{i=1}^{n-1} \frac{k}{2} \left( x_i - x_{i+1} - (-1)^i2\delta) \right)^2 ~.
\end{multline}
where $k^*:=k_h/k_v$ is the ratio between the horizontal and vertical spring constants.

We obtain equilibrium positions of $\{x_i\}$ given a strain $\vec{\varepsilon}$ by finding the roots of the forces $F_i=\partial_{x_i}E$ using the Newton Raphson method.
Each strain-driving step is divided into several small steps of $10^{-5}$ to mimic the quasi-static dynamics.
Together, the numerical model allows for a systematic in-dept study of the path-dependent behavior of our model system.

\subsection{Remarks on Spring Model} \label{sec:remarksmodel}
In the lowest order, the energy simplifies to:
\begin{equation}
	E = \sum_{i=1}^{n} \left( \frac{x_i^2}{2} - \gamma_i \right)^2 + \sum_{i=1}^{n-1} \frac{k}{2} \left( x_i - x_{i+1} - 2\delta (-1)^i \right)^2 ~,
\end{equation}
where $\gamma_i := \varepsilon_i + \delta^2/2 - \varepsilon_i^2/2 \approx \varepsilon_i + \delta^2/2$.
Hence, our model system consists of coupled quartic potentials and thus represents a general class of multistable (disordered) (meta)materials.

Additionally, in the small strain limit ($\varepsilon_i^2\approx0$), the lowest order approximation of the energy $E$ can be rescaled such that the bias $\delta$ drops out:
\begin{align}
	x_i &\rightarrow \delta x_i'~, \\
	\varepsilon_i &\rightarrow \delta^2 \varepsilon_i'~, \\
	k &\rightarrow \delta^2 k'~, \\	
	E &\rightarrow \delta^4 E'~,
\end{align}
resulting in:
\begin{multline}
	E' = \sum_{i=1}^{n} \left( \frac{x_i'^2}{2} - \varepsilon'_i - \frac{1}{2} \right)^2 + \\ \sum_{i=1}^{n-1} \frac{k'}{2} \left( x'_i - x'_{i+1} - 2 (-1)^i \right)^2 ~.
	\label{eq:energylowestorder}
\end{multline}
Although $\delta$ does not drop out in higher order approximations of the energy or for large strains, the lowest order approximation demonstrates that effective coupling strength $k$ determines the path dependent response of the model system.

\subsubsection{Critical strains}
The minimal spring model (Eq.~\ref{eq:energylowestorder}) analytically describes the evolution of the strain maps for a two-unit system. There are two critical strains: first,  
the strain $\varepsilon_\mathrm{s}$ is the strain applied at unit 2 that 
flips the orientation of unit $1$, and second, 
the pitchfork bifurcation strain $\varepsilon_{\mathrm{pf}}$.
The butterfly singularity occurs when $\varepsilon_\mathrm{s} = \varepsilon_{\mathrm{pf}}$; here the pitchfork bifurcation changes from supercritical to subcritical.

To determine $\varepsilon_\mathrm{s}$, we note at the flipping point of unit 1, the x-component of the force on unit $1$ must be zero.
Force balance then dictates that the x-component of the force of the horizontal spring is zero, so that the  x-component of the force
on unit 2 also must be zero. Therefore, the vertical springs of unit 2
must be at their rest lengths, and $\varepsilon_\mathrm{s}$ is given by geometric parameters through
the Pythagorian theorem.
Hence, $\varepsilon_\mathrm{s}$ is independent of $k$ and only depends on  $\delta$:
\begin{align}
	\left( 1 - \varepsilon_\mathrm{s} \right)^2 - \left( 2\delta \right)^2 &=  1 + \delta^2 \Rightarrow\\
	\varepsilon_\mathrm{s}  &= 1 - \sqrt{1 - 3\delta^2} \approx \frac{3}{2}\delta^2 ~.
\end{align}
As discussed above, we can rescale out the dependence on $\delta$, and so $\varepsilon'_\mathrm{s}\approx 3/2$.
We finally note that the cusps $Q_1$ and $Q_2$,
precisely are located along the lines where $\varepsilon_1=
\varepsilon_\mathrm{s}$ or $\varepsilon_2=
\varepsilon_\mathrm{s}$ (Fig.~\ref{fig:CouplingRegimes}b,c). This is because $\varepsilon_\mathrm{s}$ is the threshold between monostable and bistable systems.

\begin{figure}[t!]
	\centering
	\includegraphics[]{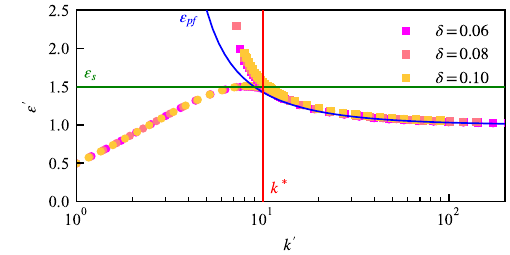}
	\caption{Evolution of bifurcation points with $k$. 
    Numerical estimate of $\varepsilon_\mathrm{pf}$ (squares) 
    and saddle node bifurcation associated with the subcritical pitchfork bifurcation (circles). The data indicate the emergence of the butterfly singularity at $k^*\approx 10\delta^2$ (red line), while the analytical approximation for $\varepsilon_\mathrm{s}$ (green line) and $\varepsilon_\mathrm{pf}$ (blue line) yield $k^*\approx9\delta^2$.}
	\label{fig:k*c}
\end{figure}

Next, we estimate the critical strain of the pitchfork singularity $\varepsilon'_\mathrm{pf}$ based on 
the lowest order approximation of the energy $E'$ (Eq.~\ref{eq:energylowestorder}). The pitchfork bifurcation occurs when the symmetric configuration at $x'_1=-x'_2$ becomes unstable, and this happens along the line $\varepsilon'_1=\varepsilon'_2$. We thus 
set $\varepsilon'_1=\varepsilon'_2=\varepsilon'$ and 
$x'_1=-1 - y$ and $x'_2=1 + y$, and 
obtain:
\begin{align}
	E'(y) = \frac{1}{2}\left( y^2 + 2y - 2\varepsilon'\right)^2 + 2k' y^2~.
\end{align}
The equilibrium position of the system is given by:
\begin{align}
	\partial_y E' = 2\left( y^2 + 2y - 2\varepsilon' \right)
    (y + 1) + 4k' y=0~,
\end{align}
which yields that in lowest order in $y$, $y\approx \varepsilon'/(1+k'-\varepsilon')$.
For typical values of $k$ and $\delta$, e.g., 
$k=0.09$ and $\delta=0.08$, $k' \gg 1$, and $k' \gg \varepsilon'$,
so in the remainder we set $y=\varepsilon' /k'$. We have checked (data not shown) that this is in quite good agreement with our 
numerical results.

Now that we have an expression for the real space configuration, 
we determine the pitchfork strain $\varepsilon_{\mathrm{pf}}$ where
the equilibrium changes from a node to a saddle. Using the 
Hessian, this occurs when
\begin{align}
	\frac{\partial^2 E'}{\partial x_1^{'2}} \frac{\partial^2 E'}{\partial x_2^{'2}} - \left(\frac{\partial^2 E'}{\partial x'_1 \partial x'_2} \right)^2 = 0~.
\end{align}
After some algebra, we find 
\begin{equation}
    \left( 3y^2 + 6y + 2 - 2\varepsilon'_{pf} + k' \right)^2 - (k')^2 = 0~,
\end{equation}
which we rewrite as
\begin{equation}
     \left( 3y^2 + 6y + 2 - 2\varepsilon_{pf}'  \right) \cdot \left( 3y^2 + 6y + 2+ 2k' -2\varepsilon_{pf}'  \right)= 0 ~.
\end{equation}

As $y$ is expected to be small, and $k'$ is of order 10, we only solve for the first term, in lowest order in $Y$ to be zero, yielding
$3y+1-1\varepsilon'=0$. Using our previous approximation, 
$y=\varepsilon'_{pf}/k'$, we finally find:
\begin{align}\label{epf}
	\varepsilon'_{pf} = \frac{1}{1 - {3}/{k'}}~.
\end{align}
As the butterfly singularity occurs when $\varepsilon'_s=\varepsilon'_{pf}$, this implies that 
it occurs when	$3/2 = {1}/({1 - {3}/{k'}})$, i.e., for $k'^*\approx 9$.
    
To check the validity of this result, we have numerically
determined $\varepsilon_\mathrm{pf}$ as function of $k/\delta^2$, and find good agreement with our approximation
(Eq.~\ref{epf}) - see Fig.~\ref{fig:k*c}. In the regime where the pitchfork is subcritical, we have also determined the corresponding saddle-node bifurcations, and we find that these reach $\varepsilon_\mathrm{pf}$ at the butterfly singularity at $k/\delta^2\approx0$ (Fig.~\ref{fig:k*c}).

\end{document}